\documentclass[12pt]{iopart}

\usepackage{graphicx}
\usepackage{amsfonts}
\usepackage{txfonts}
\usepackage{times}
\usepackage{subfigure}

\usepackage{comment}

\usepackage{textcomp}
\usepackage[font=small,labelfont=bf]{caption}
\usepackage{color}
\definecolor{gray}{gray}{0.8}

\graphicspath{{./Figures/}} 

\newcommand{\argmin}{arg\,min}
\newcommand{\beq}{\begin{equation}}     \newcommand{\eeq}{\end{equation}}
\newcommand{\beqa}{\begin{eqnarray}}    \newcommand{\eeqa}{\end{eqnarray}}
\newcommand{\bde}{\begin{description}}  \newcommand{\ede}{\end{description}}
\newcommand{\ben}{\begin{enumerate}}    \newcommand{\een}{\end{enumerate}}

\begin{document}
\title{Shortest node-disjoint paths on random graphs}

 \author{C De Bacco$^1$, S Franz$^1$, D Saad$^2$ and C H  Yeung$^{2, 3, 4}$}
\address{$^1$ LPTMS, Centre National de la Recherche Scientifique et Universite Paris-Sud 11, 91405 Orsay Cedex, France.}
 \address{$^2$ The Nonlinearity and Complexity Research Group, Aston University, Birmingham B4 7ET, United Kingdom.}
 \address{$^3$ Department of Physics, The Hong Kong University of Science and Technology, Hong Kong.}
 \address{$^4$ Department of Science and Environmental Studies, The Hong Kong Institute of Education, Hong Kong.}
\ead{caterina.de-bacco@lptms.u-psud.fr}

\begin{abstract}
A localized method to distribute paths on random graphs is devised, aimed at finding the shortest paths between given source/destination pairs while avoiding path overlaps at nodes. We propose a method based on message-passing techniques to process global information and distribute paths optimally. Statistical properties such as scaling with system size and number of paths, average path-length and the transition to the frustrated regime are analyzed. The performance of the suggested algorithm is evaluated through a comparison against a greedy algorithm.
\end{abstract}

\maketitle

\section{Introduction}

Among the various computationally-hard constraint satisfaction problems, routing and path optimization have attracted particular attention in recent years due to their non-localized nature and interdisciplinary relevance. The node-disjoint path (NDP) problem on graphs studied here, aims at finding a set of paths linking specified pairs of nodes (communications) such that no two paths share a node; the problem is classified among the NP-complete class~\cite{karp} of hard combinatorial problems. This has not only been studied as a purely theoretical problem by mathematicians in the series of graph minors~\cite{gminors} under the name of subgraph homeomorphism problem, but also by practitioners due to its wide applicability to various fields. For instance, in communication systems where the network performance is often strictly related to capacity limits, traffic congestion and the rate of information flow; and in problems of virtual circuit routing where switches located at nodes may become bottlenecks. Moreover, due to their distributive nature NDP is more resilient to failure and represents one aspect of optimal routing where network robustness is the main objective.

One specific communication application where efficient and effective NDP algorithms are essential is in the area of optical networks where transmissions using the same wavelength cannot share the same edge or vertex, hence all communications of the same wavelength must be non-overlapping (disjoint). Consequently, such an algorithm impacts on the achievable network capacity and transmission rate. In this field of routing and wavelength assignment~\cite{RWAreview}, the objective is to find a routing assignment that minimizes the number of wavelengths used. Different techniques that exploit disjoint paths heuristic algorithms have been proposed to tackle this problem; for instance, greedy algorithms~\cite{chen96,BGA}, approximations based on rounding integer linear programming formulations~\cite{banerjee1996,packing}, post-optimization methods~\cite{post14}, bin packing algorithms~\cite{skorin07}, various heuristic genetic algorithms such as ant colony optimization~\cite{blesa04} and differential evolution~\cite{DE}.

Another important application of NDP is in the design of very large system integrated circuits (VLSI), where one searches for non-overlapping wired paths to connect different integrated hardware components, to avoid cross-path interference. \\
Similarly, in wireless ad-hoc communication networks~\cite{sumpter,srinivas03,adhoc}, where each node can act as a router, path overlaps imply signal interference and low transmission quality, whereas longer paths imply poor signal to noise ratio due to multiple relays and higher transmission power; hence the need to consider both path length and transmission overlaps to be minimized is essential for routing problems. Solutions to the NDP problem also provide fault tolerant routes due to the optimal separation of communication paths all over the network, so that if a node (router) fails, as frequently happens in wireless networks due to the mobility of hosts, only few communications will be affected~\cite{Li2004,jain2008}. This feature is particularly important when quality of service (QoS) is one of the main requirements in the set up of a communication network, along with the load-balancing feature of NDP that prevents network congestion by establishing non-overlapping routes. This is especially relevant to connection-oriented networks~\cite{CN} that are strongly affected by node failures and congestion~\cite{qos2006}.  \\

Practical algorithms for various applications often depend on the specific network topologies considered~\cite{mesh96} and mostly focus on the optimization version of the problem, i.e. maximizing the number of paths routed~\cite{mndp}. The satisfiability version of the problem, i.e. whether all paths can be routed successfully without overlap, is not considered; theoretical studies often give bounds to the achievable approximation instead of providing a practical algorithm for individual instances and fail to calculate path lengths and possible overlaps at the same time as part of the optimization process observables.
Given that paths are constrained to be contiguous and interaction between paths is non-localized, a local protocol is insufficient and global optimization is required. The computational complexity is determined by the fact that such a global optimization problem has to consider all variables simultaneously in order to minimize a cost function with non local interactions between variables.

Unlike other constrained satisfaction problems on networks, NDP has received little attention within the statistical physics community. In this paper we consider a random version of NDP on regular graphs (Reg),  Erd\H{o}s R\'{e}nyi (ER)~\cite{ER}  and a dedicated type of random graph (RER) described in Section \ref{sec:results}, with the aim of testing the efficacy of statistical physics-based methods derived in the context of spin glass theory~\cite{sg87} such as
belief propagation or message-passing (MP) cavity method~\cite{inf09,cavity} as viable alternatives to greedy algorithms; we also study statistical and scaling properties of quantities of interest as a function of network size and number of paths. We study sparse regular, ER and RER random graphs as they are the most interesting for the problem at hand, but the methodology can be easily extended to accommodate other sparsely connected architectures. Clearly, due to the hard constraint of node disjoint paths, typically no solutions would be found in graphs having a non negligible number of nodes with degree $k=1,2$. Moreover, graphs with a small number of high degree nodes (hubs) or with high modularity measure, such as scale-free or planar graphs, are not interesting for the node-disjoint routing problem since when a paths passes through one of these special nodes it leads directly to graph fragmentation, hence frustration. The situation would be very different for constraints on edges instead, but this variant of the problem is left for future work. Finally, the requirement for the graph to be sparse is suggested by restrictions on the validity of the cavity method which is based on fast decaying correlation functions, i.e. a negligible number of loops in the graph.

Numerical simulations indicate that MP outperforms greedy breadth-first search algorithms not only in finding better solution but also in reaching a higher frustration threshold. Moreover, we find scaling of the expected total length of the NDP as a function of the system size and graph connectivity that goes as $\frac{M\log V}{V\log^{\gamma}(k-1)}$ with exponent $\gamma$ that depends on the type of graph, where $V$ is the number of nodes and $M$ the number of paths. We find good agreements between theory and simulation data for graphs of average degrees $k=3,5,7$ and sizes $V=1000, 2000, 4000, 5000, 10000$. Finally, we study statistical properties of physical quantities observed a posteriori, i.e. when a solution is found, such as path length distribution, degree distribution and maximum cluster size for the case of regular graphs.

The reminder of the paper is organized as follows: in Section~\ref{sec:model} we will introduce the model used followed by the algorithmic solution in Section~\ref{sec:solution}. Results obtained from numerical studies will be presented in Section~\ref{sec:results} followed by conclusions and future research directions in Section~\ref{sec:conclusion}.

\section{Model}
\label{sec:model}
Given an undirected graph (or network) $\mathcal{G}=(\mathcal{V},\mathcal{E})$ characterized by $V=|\mathcal{V}|$ nodes and $E=|\mathcal{E}|$ edges we define a set of $M$ communications $\mathcal{C}$ as paths on edges of the graph, each of which  originates from a source node $S$ and terminates in a receiver node $R$. We introduce a variable $\Lambda ^{\mu}_{i}$ to characterize each node $i \in \mathcal{V}$:
\begin{equation}
\Lambda^{\mu}_{i}= \left\{  \begin{array}{cl}
 +1 & \mbox{ if $i$ is a sender for communication $\mu$} \\
 -1 & \mbox{ if $i$ is a receiver for communication $\mu$}  \\
0 &  \mbox{ if $i$ is neither a sender nor a receiver for communication $\mu$ }
\end{array} \right.
\end{equation}
The full node characterization is specified by a vector $\bar{\Lambda}_{i}:=(\Lambda^{1}_{i},\dots, \Lambda^{M}_{i})$ of modulus $||\bar{\Lambda}_{i}||:=\sum_{\mu=1}^{M} |\Lambda_{i}^{\mu} | \in \{0,1\}$, where $ |\Lambda_{i}^{\mu} |$ denotes the absolute value of $\Lambda_{i}^{\mu}$; $||\bar{\Lambda}_{i}||$ is $0$ if $i$ is neither a sender nor a receiver for any communication, termed a transit node, and $1$ if $i$ is either a sender or a receiver of some communication. In this way each node can send or receive at most one communication.

\begin{figure}[htbp]
\begin{center}
  \includegraphics[width=10cm]{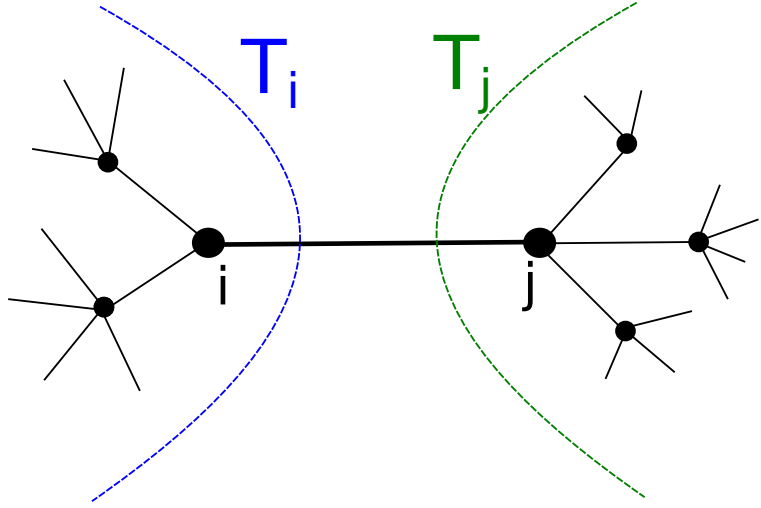}
  \caption{Subtree division. $T_{i}$ is the subtree rooted in $i$ when the edge $(ij)$ is removed. Conversely, $T_{j}$ is the subtree rooted in j after the same edge removal. }
\label{tree}
\end{center}
\end{figure}

For a given set of $M$ sender-receiver pairs $( S^{\mu}, R^{\mu})$ with $\mu=1, \dots,M$ we address the problem of finding a set of communications that optimize a cost function which penalizes path length and prevents communications overlap (traffic). The state of the network can be specified by introducing a variable $I^{\mu}_{ij}$ for each edge $(ij) \in \mathcal{E}$ and for each communication $\mu$, which specifies whether communication $\mu$ passes through edge $(ij)$ and in which direction:
\begin{equation}
I^{\mu}_{ij}= \left\{  \begin{array}{cl}
 +1 & \mbox{ if $\mu$ passes through $(ij)$ from $i$ to $j$} \\
 -1 & \mbox{ if $\mu$ passes through $(ij)$ from $j$ to $i$}  \\
0 &  \mbox{ if $\mu$ does not pass through $(ij)$ }
\end{array} \right.
\end{equation}
Notice that in this formalism $I^{\mu}_{ij}=-I^{\mu}_{ji}$. We term these variables currents and define for each edge $(ij)$ a vector $\bar{I}_{ij}:=(I^{1}_{ij},\dots, I^{M}_{ij})$ that collects information on all currents involved in that edge. 	
Currents are subject to Kirchhoff law:
\begin{equation}\label{constraint}
\sum_{j \, \in \partial \, i} I^{\mu}_{ij}- \Lambda^{\mu}_{i}=0 \qquad \forall \mu=1,\dots, M~.
\end{equation}
For a given path optimization problem we seek the communication configuration $\mathcal{C}^{*}$ that minimizes a cost function $c(\{ \bar{I}_{ij} \} )$, which penalizes path length and traffic congestion:
\begin{equation}\label{cost}
c(\{ \bar{I}_{ij} \})=\sum_{(ij)\in \mathcal{E}} f(||\bar{I}_{ij}||)
\end{equation}
where $ f(||\bar{I}_{ij}||)$ is a monotonically increasing function of $||\bar{I}_{ij}||:=\sum_{\mu=1}^{M}|I_{ij}^{\mu}| $, that penalizes both congestion and path length; where $|I_{ij}^{\mu}|$ denotes the absolute value of $I_{ij}^{\mu}$.\\
We would like now to search for approximate solutions to this problem by message-passing equations~\cite{inf09}. To derive a distributed algorithm it is useful first to consider tree-like graphs $\mathcal{T}$, for which one can derive exact recursive equations, and later on use these equations as an approximation for arbitrary graphs $\mathcal{G}$.\\
If $\mathcal{T}$ is a tree, the removal of any edge $(ij) \in \mathcal{E}$ divides $\mathcal{T}$ in two disjoint subtrees $T_{i}$ and $T_{j}$ (see figure~\ref{tree}).
We can define $\hat{E}_{ij}(\bar{I})$ as the optimized cost on $T_{i}$ when the current $\bar{I}$ flows through on the edge $(ij)$; in this way we can write the message sent from node $i$ to his neighbor $j$ when the current $\bar{I}_{ij}$ flows through the edge $(ij)$ as $E_{ij}(\bar{I}_{ij}):=\hat{E}_{ij}(\bar{I}_{ij}) + f(||\bar{I}_{ij}||)$.

The messages $E_{ij}(\bar{I}_{ij} )$ admit the \textit{min-sum}~\cite{inf09} recursion relation:
\begin{equation}\label{iterate}
E_{ij}(\bar{I}_{ij})=\min_{\bar{I}_{ki}| \small{constraint}} \left\{ \sum_{k \in \partial i \backslash j } {E_{ki} (\bar{I}_{ki})} \right\}+ f(||\bar{I}_{ij}||) ~,
\end{equation}
where the symbol $\partial i$ stands for the set of neighbors of node $i$ and the \textit{constraint} is the Kirchoff law (\ref{constraint}).\\
In the following we will use the recursion equation~(\ref{iterate}) on arbitrary random graphs to approximate the constrained minimum of $c(\{\bar{I}_{ij}\})$, the cost defined in equation~(\ref{cost}). Namely:
\begin{equation}
c^{*}:=\min_{\bar{I}} \frac{1}{E}\sum_{(ij)\in \mathcal{E}} \left\{ E_{ij}(\bar{I})+E_{ji}(-\bar{I}) -f(||\bar{I}||)  \right\}
\end{equation}
where the last subtracted term is introduced to avoid double counting the cost of edge $(ij)$.
\\
Unfortunately, the computational complexity of this algorithm is exponential in the number of communications $M$. In fact, messages $E_{ij}(\bar{I}_{ij})$ can a priori take $3^{M}$ values corresponding to all possible currents passing through a single edge $(ij)$. Therefore, we cannot generally treat even moderately large values of $M$~\cite{pnas13}. The problem can be simplified if we introduce the hard constraint that paths cannot overlap on nodes (and thus neither on edges). This has the important consequence of reducing the configuration space from $3^{M}$ to $2M+1$ and the computational complexity becomes linear in $M$. This restricted version of the path optimization problem is called the node-disjoint path problem (NDP), as we already mentioned in the introduction, and is the problem we address here. Notice that since we impose the node-disjoint constraint for the communications, then one communication at most flows through the edges, so that $||\bar{I}_{ij}||=\sum_{\mu=1}^{M}|I_{ij}^{\mu}|\in\{0,1\}.$ This corresponds to taking:
 \begin{equation}
 f(||\bar{I}||)=\left\{ \begin{array}{ll}
 \infty & \mbox{if} \quad ||\bar{I}|| \geq 2 \\
 1 &  \mbox{if} \quad ||\bar{I}||=1\\
  0 & \mbox{if} \quad  ||\bar{I}|| =0
 \end{array}  \right.
 \end{equation}
so that the cost function~(\ref{cost}) represents indeed the total path length. \\
In order to solve equation~(\ref{iterate}) iteratively we define a protocol for taking into account only the allowed configurations at each edge given the current value $\bar{I}$ passing through it and $\bar{\Lambda}_{i}$ at vertex $i$.

If $| \bar{\Lambda}_{i}|=0$ then:
\begin{eqnarray} \label{Izero}
E_{il} (\bar{I}_{il}=\bar{0})&=&\min\left\{ \sum_{j \in \partial i \backslash l } {E_{ji} (\bar{I}_{ji}=\bar{0})}, \right.\\
&\phantom{=}& \hspace{-15mm} \left. \min_{j_1, j_2 \in \partial i \backslash l;\mu \in M} \left[ E_{j_1 i} (I_{j_1 i}^\mu =+1) + E_{j_2 i} (I_{j_2 i}^\mu =-1) + \sum_{k \in \partial i \backslash l, j_1,j_2 } {E_{ki} (\bar{I}_{ji}=\bar{0}) } \right]  \right\} \nonumber
\end{eqnarray}

\begin{eqnarray} \label{Ione}
E_{il} (I_{il}^{\mu}=\pm 1)&=& \min_{j \in \partial i \backslash l } \left \{ E_{ji} (I_{ji}^{\mu}=\pm 1)
+  \sum_{k \in \partial i \backslash l,j } E_{ki} (\bar{I}_{ki}=\bar{0}) \right\} +1
\end{eqnarray}

If $ \Lambda_{i}^{\mu}=\pm 1$ then:
\begin{eqnarray}\label{SR}
E_{il} (\bar{I}_{il}=\bar{0})&=& \min_{j \in \partial i \backslash l } \left \{ E_{ji} (I_{ji}^{\mu}=\mp 1)
+  \sum_{k \in \partial i \backslash l,j } E_{ki} (\bar{I}_{ki}=\bar{0}) \right\} \\
E_{ji} (I_{ji}^{\nu} =\pm 1) &=& +\infty \quad (\nu \neq \mu)\\
E_{ji} (I_{ji}^{\mu}=\mp 1) &=& +\infty \\
E_{ji} (I_{ji}^{\mu}=\pm 1) &=& \sum_{j \in \partial i \backslash l } {E_{ji} (\bar{0})} + 1 \label{SRend}
\end{eqnarray}
The constant $+1$ that appears equations~(\ref{Ione}) and (\ref{SRend}) are the costs assigned for a unit of current passing through the considered edge (i.e. $f(1)=1$). This cost is the one required for the shortest paths but can be generalized to other arbitrary types of costs.

\begin{figure}[htbp]
\begin{center}
  \includegraphics[width=16cm]{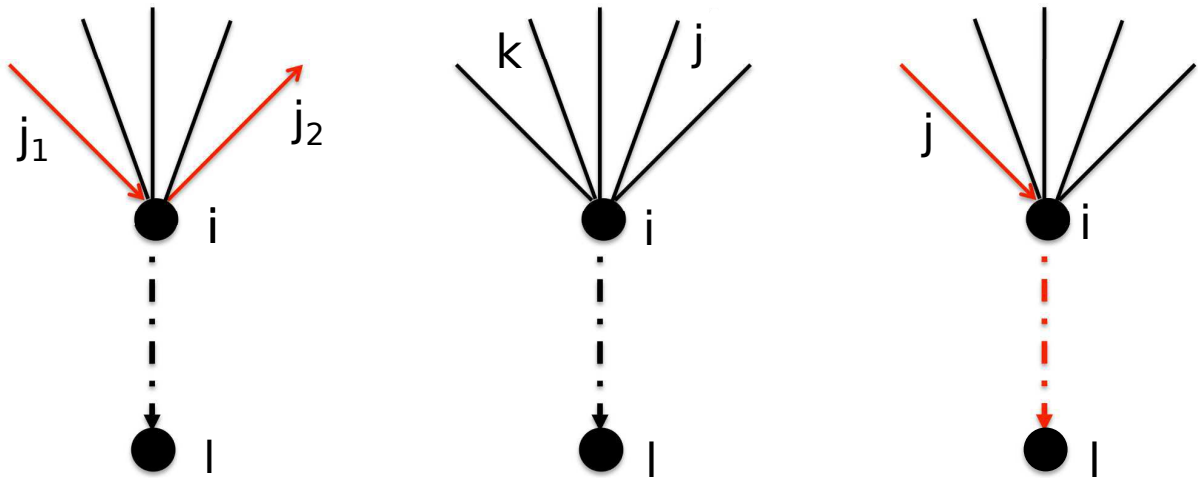}
  \caption{Transit node cavity diagram. Left and center represent the two terms inside the $\min$ brackets in equation ~(\ref{Izero}). Right represents equation ~(\ref{Ione}) . }
\label{diagram}
\end{center}
\end{figure}

Equation~(\ref{Izero}) represents the case where $i$ is a transit node and no current passes through edge $(ij)$, then the allowed configurations are that either no currents pass through the remaining neighboring edges (first term inside curly brackets) or one current enters and then exits $i$ through a pair of neighboring edges, all others edges being unused (second term inside brackets). In figure \ref{diagram} you can see a diagram representing the different allowed configurations for a transit node. Equation~(\ref{Ione}) represents the case where $i$ is a transit node and the communication $\mu$ passes through edge $(ij)$; in this case the only allowed configuration is that where the same communication $\mu$ enters/exits from one of the other neighboring edges, all others being unused. Similar considerations are used to formulate the equations~(\ref{SR}-\ref{SRend}) for senders and receivers.

The procedures of applying the algorithm can be summarized as follows:
\begin{itemize}
\item Initialize messages at random.
\item Pick in random order all $i \in \mathcal{V}$ and update messages using (\ref{Izero}), (\ref{Ione}) and (\ref{SR}-\ref{SRend}) until convergence is reached (i.e., message changes are below a given threshold).
\item Use the converged messages to calculate physical observables.
\end{itemize}


\begin{figure}[htbp]
\begin{center}
  \begin{tabular}{cc}
  \includegraphics[width=8cm]{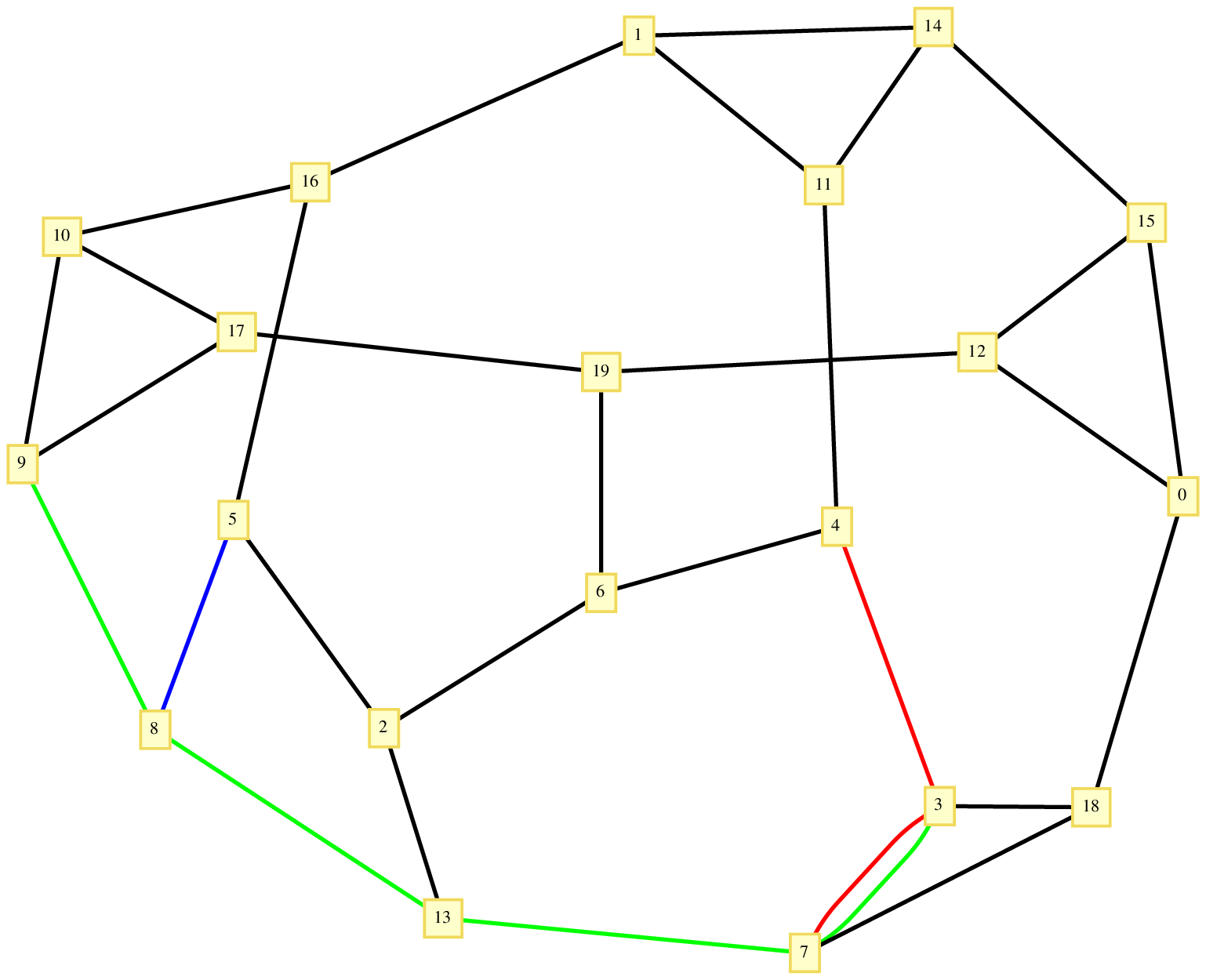} &
\includegraphics[width=8cm]{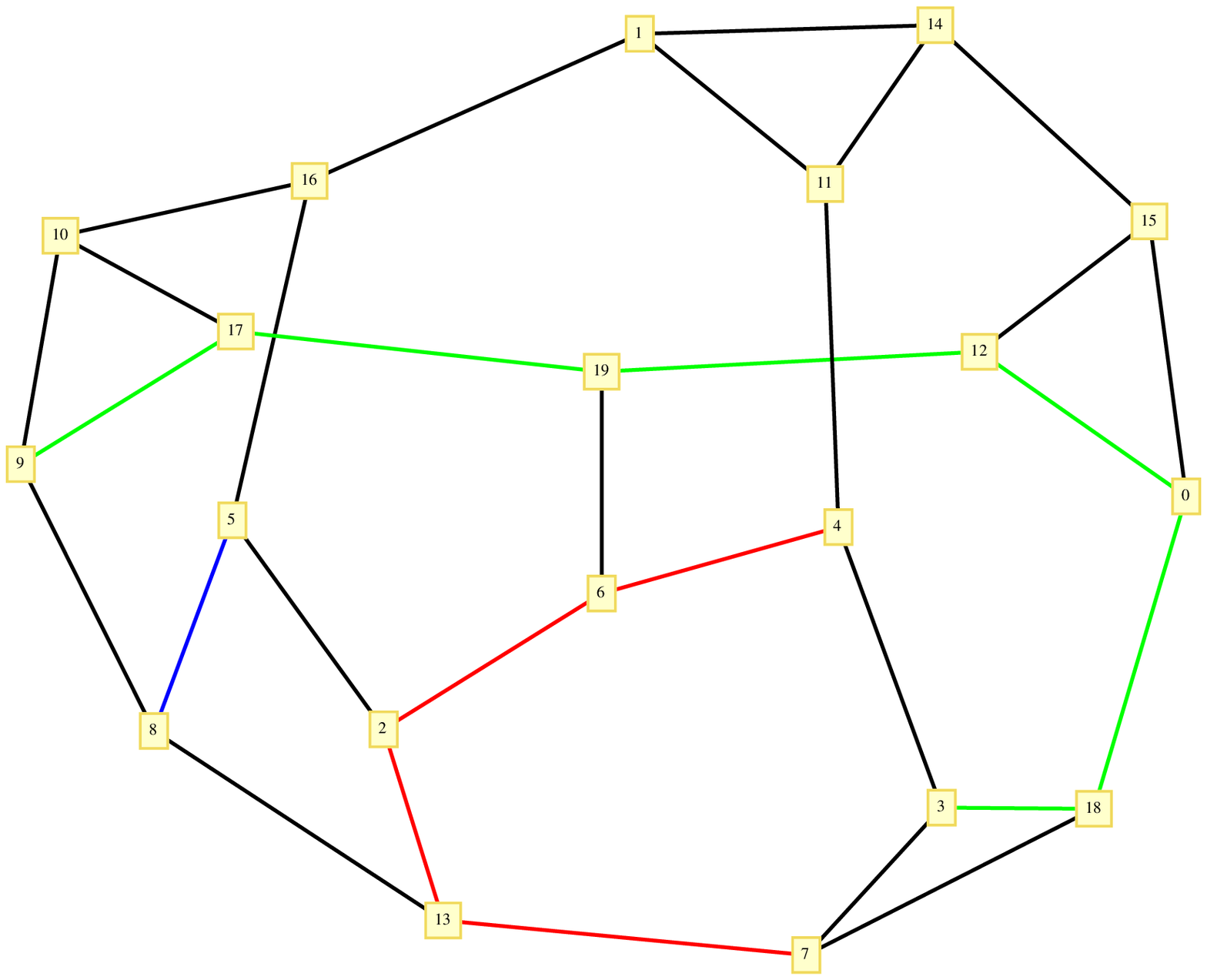}
  \end{tabular}
  \caption{Edge-disjoint routes. For a single instance of a graph of size $V=20$ and $M=3$ communications we have on the left the shortest paths and on the right the optimal non-overlapping solutions. We can see that the green path has to be re-routed to avoid both blue and red communications. Also the red one cannot take its shortest path because the sender of the green communication is on that path.}
\label{graph}
\end{center}
\end{figure}

\section{Obtaining a solution}
\label{sec:solution}
Once the iterative equations (\ref{Izero}), (\ref{Ione}) and (\ref{SR}-\ref{SRend}) have converged, the resulting messages can be used to calculate the solution. We define the energy per link~\cite{cavity}:
\begin{equation}
E_{ij}^{Link}(\bar{I}):=\left\{E_{ij}(\bar{I})+E_{ji}(-\bar{I}) - ||\bar{I}|| \right\}
\end{equation}
where the last term on the right $ ||\bar{I}|| $ is subtracted to avoid double counting as it appears in both of the previous two terms. To find a solution we calculate:
\begin{equation}\
E_{ij}^{*Link}:=\min_{\bar{I}} E_{ij}^{Link}(\bar{I})
\end{equation}
 for each link in the graph and store the current values that minimize the energy per link for each edge:
 \begin{equation}
\bar{I}_{ij}^{*}:=\argmin_{\bar{I}} ~{E_{ij}^{Link}(\bar{I})}~.
\end{equation}
Eventually, we sum over all $(ij) \in \mathcal{E}$ to find the different paths and total length:
 \begin{equation}
L_{tot}:=\sum_{(ij) \in \mathcal E}{||\bar{I}_{ij}^{*}||}~.
\end{equation}
In the cavity formalism \cite{cavity} this is equivalent to calculating the quantity:
\begin{equation}
E_{i}:=\min_{\bar{I}_{ki} | constraint}{\sum_{k \in \partial i}{E_{ki}(\bar{I}_{ki})}}~,
\end{equation}
which represents the energy per node. Finally, the total energy (or path length) is the combination of the two, which in the case of a $k-$regular graph establishes the formal relation:
\begin{equation}
E_{t}=\sum_{i \in \mathcal{V}}E_{i} -\frac{k}{2}\sum_{(ij) \in \mathcal{E}}E_{ij}^{Link}~.
\end{equation}

Notice that the calculation of $E_{ij}^{Link*}$ is carried out link by link
as if the energies per link were statistically independent. It is not intuitively clear that doing this will result in the optimal paths which do not overlap and are also fully connected from the source to the receiver. This is a consequence of having used messages which implicitly contain global information on the constraints and path lengths, so that the energies per link are indeed  globally interdependent albeit in a non-obvious manner.

To fully characterize the solutions statistically we calculate also other observables as explained below. Finally, we calculate the paths and corresponding lengths in a sequential order using a greedy breadth first-search local algorithm (BFS) to compare the results obtained against our MP-based algorithm.

\subsection{Algorithmic complexity}
The node-disjoint constraint is very restrictive and algorithmically helpful in comparison to other routing models where overlaps are allowed but minimized~\cite{pnas13,competition12}. This hard constraint is indeed paramount in reducing considerably the algorithm's computational complexity. If we allow for overlaps we need to span a configuration space of the order of $3^{M}$ at each cavity iteration, leading to a complexity of $O( N\,(3^{M})^{k-1})$, where the exponent $k-1$ explores the different flux combinations for each of the $k-1$ independent neighboring sites of the considered message; there are order of $N$ such messages. In order to tackle this issue proper approximations have to be introduced as in~\cite{pnas13,competition12} where they use techniques from polymer physics~\cite{pnas13} or convexity properties of the cost function~\cite{competition12}. On the contrary when the overlap is prohibited we reduce the configuration space from $3^{M}$ to $(2M+1)$, as this is the number of allowed configurations (the term $2M$ is derived from the number of possible currents $I^{\mu}=\pm 1$ and the additional $+1$ is due to the configuration of all $0$), hence there is no need for approximations because the entire configuration space can be efficiently calculated by the cavity equation. Actually, the use of cavity MP implicitly requires one important approximation as it assumes that when node $i \in \mathcal{V}$ is removed, all its neighbors are statistically independent. This is equivalent of having fast decaying correlation function between these neighboring nodes. This hypothesis is verified in trees and in locally tree-like sparse graphs.

For the same reason is important to distinguish between edge and node overlaps. In this work we chose to consider constraints on nodes motivated by the reduced complexity as explained before; in case of edge constraints one has to consider a much bigger configuration space where all configurations with different communications entering and exiting the same transit node must be considered in the optimization routine. For this reason approximations should be introduced as in the case of the models which minimize overlap. The edge-disjoint variant of the problem will be left for future work.

We performed single instance simulations to find optimal microscopic solutions; to obtain macroscopic averages one would usually use population dynamics, one of the most commonly used numerical tools in statistical mechanics literatures~\cite{pd01,inf09} for studying similar models. Population dynamics is considered when the thermodynamic limit $V \rightarrow \infty$ is taken and the system size is not fixed a priori as in the single instance algorithm. In our model the use of population dynamics does not make much sense since the parameter $M$ enters explicitly in the expressions of the messages because it represents the domain of the fluxes, which is of size $2M+1$. But when we fix $M$ at the same time we are fixing a system size $V$, because we extract random pairs $(S,R)$ with density $M/V$. Hence, it is impossible to decouple the messages domain from the system size, preventing us to properly employ the thermodynamic limit through population dynamics. There is also another problem, that such a macroscopic oriented approach would introduce averages over all possible configurations $(S,R)$, including both frustrated and unfrustrated configurations with much higher energies. Thus the macroscopic averages are highly biased by the fewer frustrated configurations and more complex algorithm should be designed to discard such cases. For these two reasons we did not consider in the following the population dynamic counterpart of the algorithm but focused only on averages over single instances.

\subsection{Greedy algorithm}
To test the performance of the algorithm we compared the results obtained with those given by a greedy algorithm (or its variant) that is often used in literatures to solve the NDP problem in different contexts~\cite{sumpter, srinivas03, chen96,BGA}. The greedy protocol considers only local information around the sources and then builds up a solution step by step recursively, hence reducing considerably the complexity but at the same time completely ignoring other communication positions in the network.

A typical greedy algorithm works in the following way: start by choosing an arbitrary pair $(S,R)$, find the shortest path linking the two nodes and then remove nodes belonging to this path from the available network nodes. Choose a second pair and repeat the procedures until either all the $M$ paths from sources to destinations have been established or no solution can be found due to frustration. Clearly, the performance of this algorithm is strictly dependent on the order in which we choose the pairs. For instance, in the extreme case the first pair selected is the one with the longest shortest path among all the $M$ communications; this implies that we have effectively a more restricted graph and choice of paths, leading for a long second path and even more restricted choice of paths later on.

\section{The results}
\label{sec:results}
We performed numerical simulations on three types of random graphs. Standard regular random graphs (Reg): each node has fixed degree $k$; Erd\H{o}s R\'{e}nyi random graphs (ER) ~\cite{ER}: edges are drawn at random between each pair of nodes with probability $p = \langle k \rangle /V$; a decorated random graph (RER): starting from a regular random graph of degree $k_{1}$ (which is the minimum degree of this graphs), we then randomly add new edges as in the ER model until the final average connectivity is $\langle k \rangle = k > k_{1}$. Notice that the degree distribution in this case can be obtained by writing $k = k_{1}  +  d$ where $d$ is Poisson distributed with $\langle d \rangle = k  -  k_{1}$. The parameters used were average degrees $\langle k \rangle  = 3, 5, 7$ and system sizes of $V = 1000, 2000, 4000, 5000, 10000.$ We calculated averages over $[50-500]$ realizations for both the MP and the greedy algorithms; we used a smaller number of realizations for cases of higher complexity (as for $V=10^{4}$ and $k=7$). Nevertheless, results in all cases are stable and with small error bars with respect to the symbols used. We omitted the error bars from the figures for clarity.

\begin{figure}[htbp]
\begin{center}
\centering
\includegraphics[width=16cm]{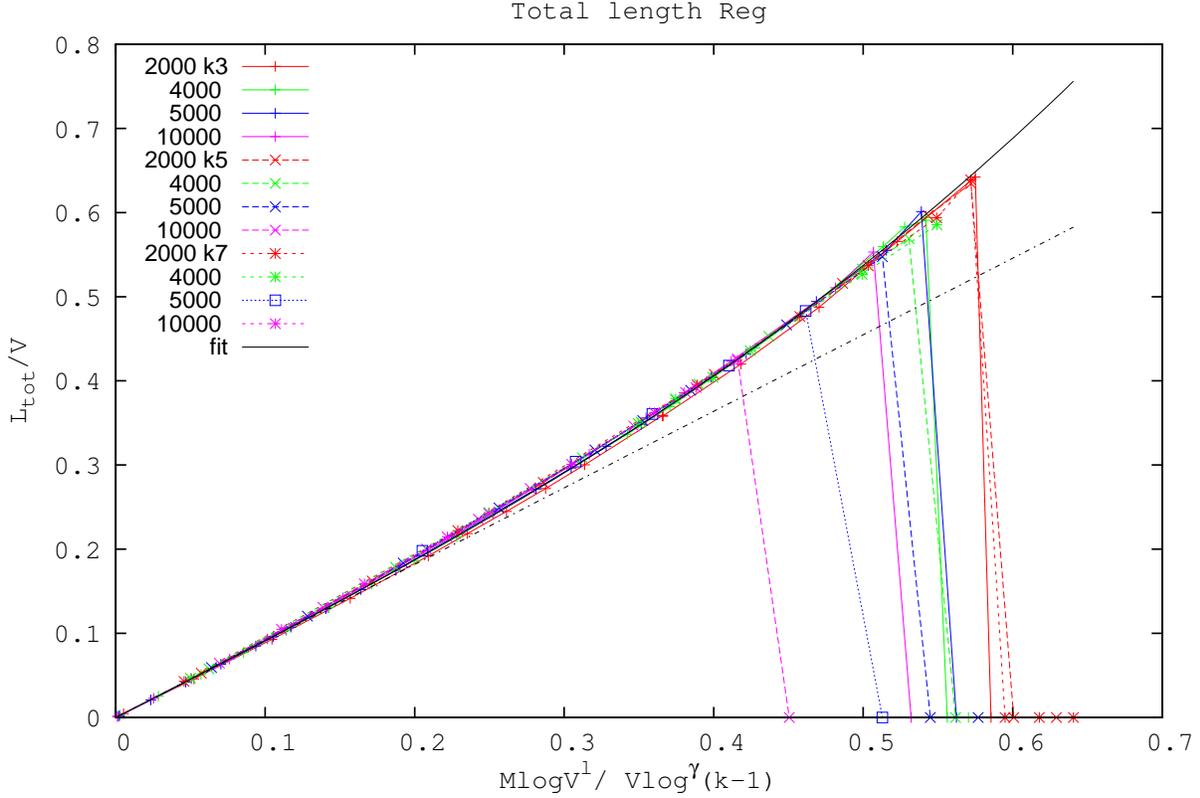}
\caption{Expected total normalized path length -  regular graphs (Reg). We obtained results for regular graphs of fixed degree $k=3,5,7$ and for system size $V=2000,4000,5000,10000$. We found a global scaling rule with respect to the variable $x:=M\log V^{l}/V \log^{\gamma}(k-1)$, with $l=1.00$ and $\gamma=0.87$. In black (solid line) we draw the cubic fit $ax+cx^{3}$, with $a=0.910 \pm 0.001$ and $c=0.660 \pm 0.009$. The dotted-dashed line represent the shortest path (not considering overlap) trivial solution. Error are not reported because smaller or comparable to point sizes.}
\label{reg_length}
\end{center}
\end{figure}

\begin{figure}[htbp]
\begin{center}
\centering
\includegraphics[width=16cm]{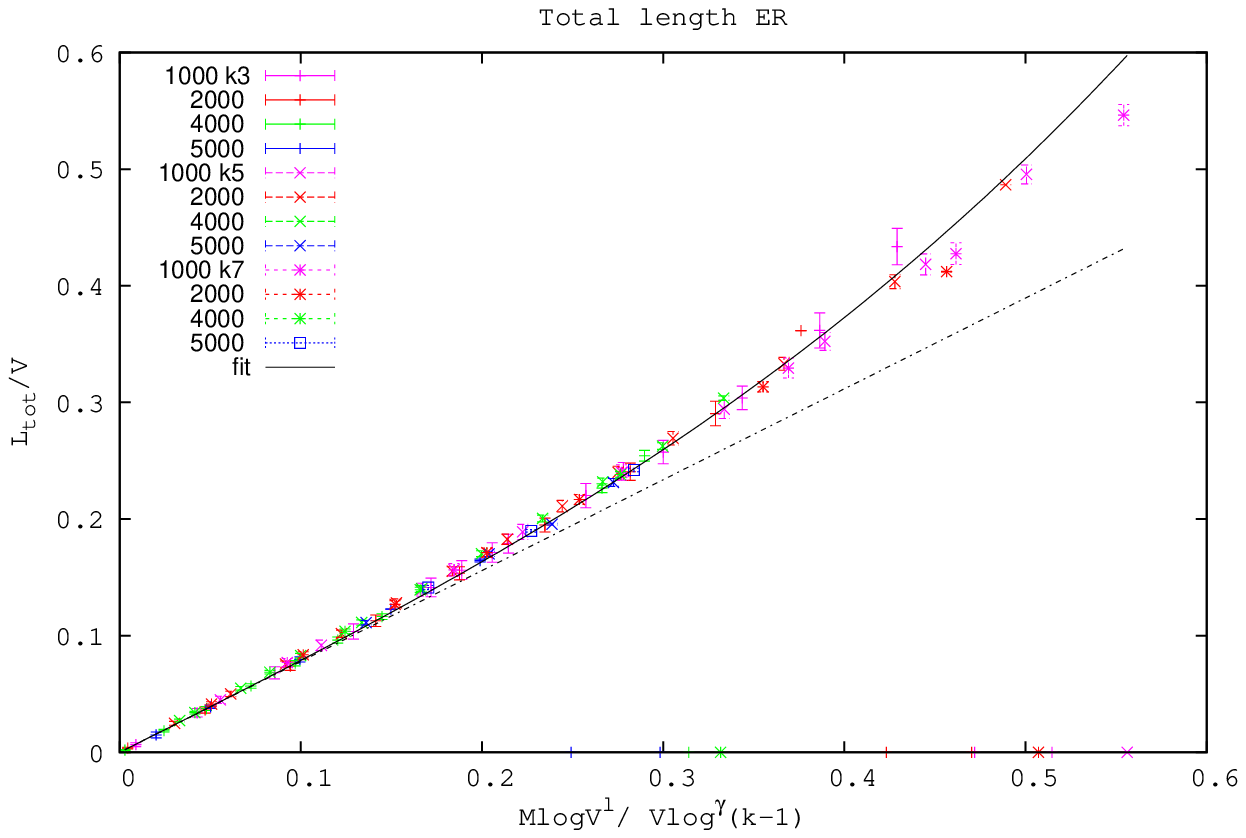}
\caption{Expected total normalized path length - Erd\H{o}s R\'{e}nyi  (ER). We obtained results for ER graphs of expected degree $\langle k \rangle=3,5,7$ and for system size $V=1000, 2000,4000,5000$. We found a global scaling rule with respect to the variable $x:=M\log V^{l}/V \log^{\gamma}(k-1)$, with $l=1.00$ and $\gamma=0.69$. In black (solid line) we draw the cubic fit $ax+cx^{3}$, with $a=0.77 \pm 0.01$ and $c=0.96 \pm 0.03$. The dotted-dashed line represent the shortest path (not considering overlap) trivial solution.}
\label{er_length}
\end{center}
\end{figure}
We found a system size scaling that is a cubic function of the variable $x:=\frac{M\log V}{V \log^{\gamma}{(k-1)}}$. A qualitative explanation of the scaling is as follows. The average path length in random graphs goes as $\langle l \rangle \sim \log V/ \log \langle k \rangle$ (see~\cite{barabasi02} for an extensive review of graphs properties) and in our case we have $M$ paths to consider.We can refine the dependence on $k$ using instead $\langle l \rangle \sim \log V/ \log{(k-1)}$. Now, suppose all communications take their shortest path, the quantity $x=M \log V/ \log^{\gamma}{(k-1)}$ would be a good estimate of graph occupancy for the NDP, where the exponent $\gamma$ has been introduced as a free parameter to account for the approximation in the expression for $\langle l \rangle$ as a function of $k$ for different types of graphs. Furthermore, if we divide by the number of available nodes $V$ we can define the occupancy ratio as $ \frac{M \log V}{ V \log^{\gamma}{(k-1)}}=x$. Therefore in this simple case we would expect $L_{tot}/V$ increasing linearly in $x$. If overlaps are prohibited, for a sufficiently high value of $M$ the communications are increasingly forced to take longer routes, leading to a faster than linear increase in the scaling variable $x$. From numerical simulations we found for the NDP a cubic increase $\frac{L_{tot}}{V}=ax+c x^{3 }$ in the scaling variable $x=\frac{M\log V}{V \log^{\gamma}{(k-1)}}$. For small $M$ this function agrees well with the linear shortest path behavior but for values of $x >0.2$ the steeper increase of $L_{tot}$ becomes predominant.
Figures ~\ref{reg_length} and ~\ref{er_length} show a good data collapse of the normalized expected total length per node $L_{tot}/V$ as a function of the scaling variable $x$ for different graph connectivities for Reg and ER graphs respectively. We notice a first regime where the curves follow the linear behavior of the dashed line representing the shortest paths. The term ``sparse regime'' is used since paths are sufficiently far apart, $M$ is small, and then no re-routing is needed as each communication will simply take its shortest path. For $x>0.2$ the curves show the cubic steeper behavior that represents the increase in path lengths to avoid overlaps. The term ``dense phase'' reflects the increase in path density; $M$ is sufficiently high so that shortest-path choices induce conflicting demands and communications are rerouted, taking longer paths to avoid overlaps.

Finally, for large $M$ we can identify different frustration points, represented by vertical lines in figure  \ref{reg_length}, that connect the largest $M$ for which solutions have been found with the points where frustration is reached and the length is set to $0$ by convention. We see that the frustration points do not collapse and that the bigger the graph size $V$ and the higher the connectivity $k$, the earlier frustration sets in (as a functions of $x$). Arguably, this is due to algorithmic convergence rather than theoretical arguments. In fact the higher $V$ and $k$ are, the higher the corresponding algorithmic complexity, and hence the larger the number of iterations required to reach convergence. Due to the prohibitive computation cost we ran a smaller number of instances for higher values of $V$ and $k$, and without increasing the preset maximum convergence time. We suspect that convergence can be reached in these cases albeit in a much longer times, and hence a solution could be found as well in theory, but has not been found due to the computational limits imposed. Hence we can not provide a precise measure for the frustration transitions nor make further statements regarding their collapses for different systems sizes and connectivities.

\begin{figure}[htbp]
\begin{center}
\centering
\includegraphics[width=16cm]{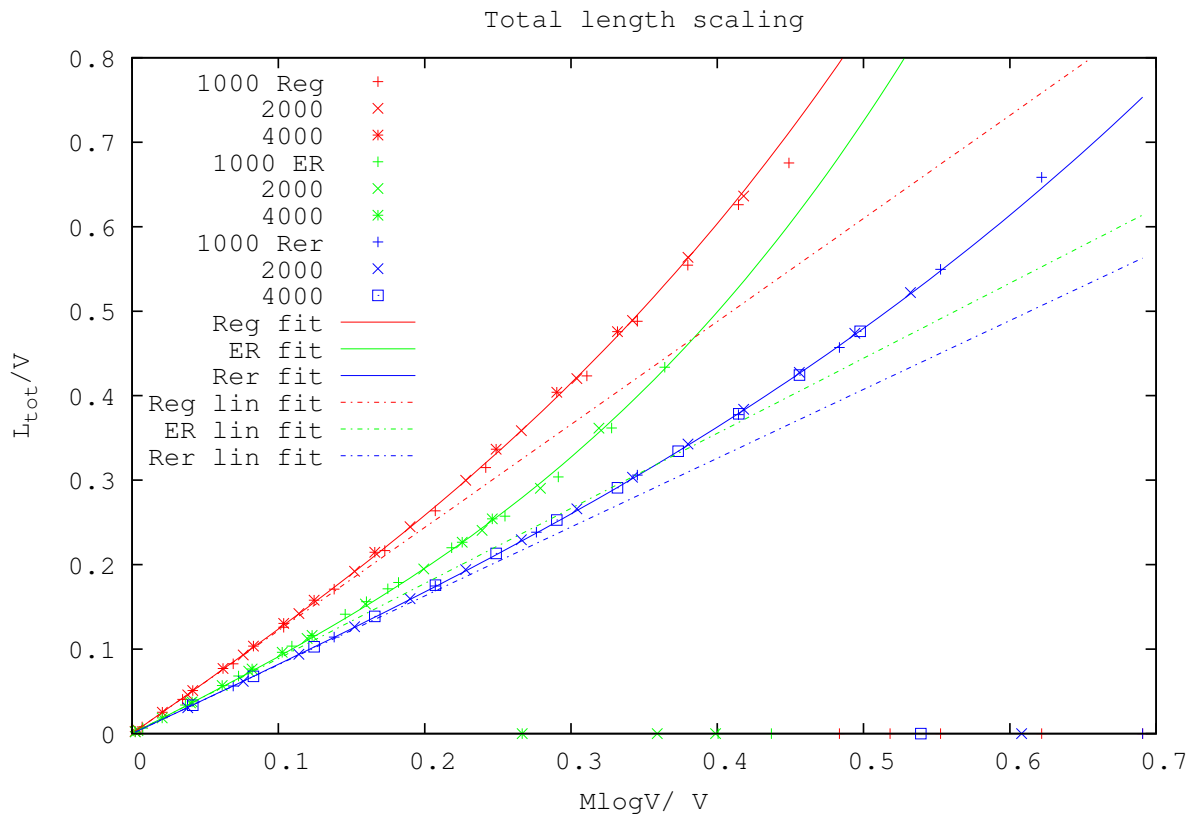}
\caption{Scaling in total length. We plotted the scaling of the total length per node as a function of $M\log V/V$ for Reg, ER and RER graphs of different system sizes; $\langle k\rangle=3$ for Reg and ER and $\langle k\rangle=4$ for RER. We can see different slopes in the cubic fits of the  various curves, for instance ER graphs achieve shorter lengths but with higher cubic slope, meaning that their length increases faster with traffic due to the smaller number of path choices. }
\label{all_length}
\end{center}
\end{figure}

In figure \ref{all_length} we can see the scaling behavior for Reg, ER and RER of given average connectivity and different system sizes; we fixed $\gamma=0$ arbitrarily to highlight the dependence on $V$. We can notice how different types of graph, although having different average lengths, follow the same cubic scaling in $x$. The steeper slope of the ER graphs shows the smaller number of paths choices in this type of graphs that forces the path to rewire in increasingly more convoluted patterns and hence also reach frustration earlier.

\begin{figure}[htbp]
\centering
\includegraphics[width=16cm]{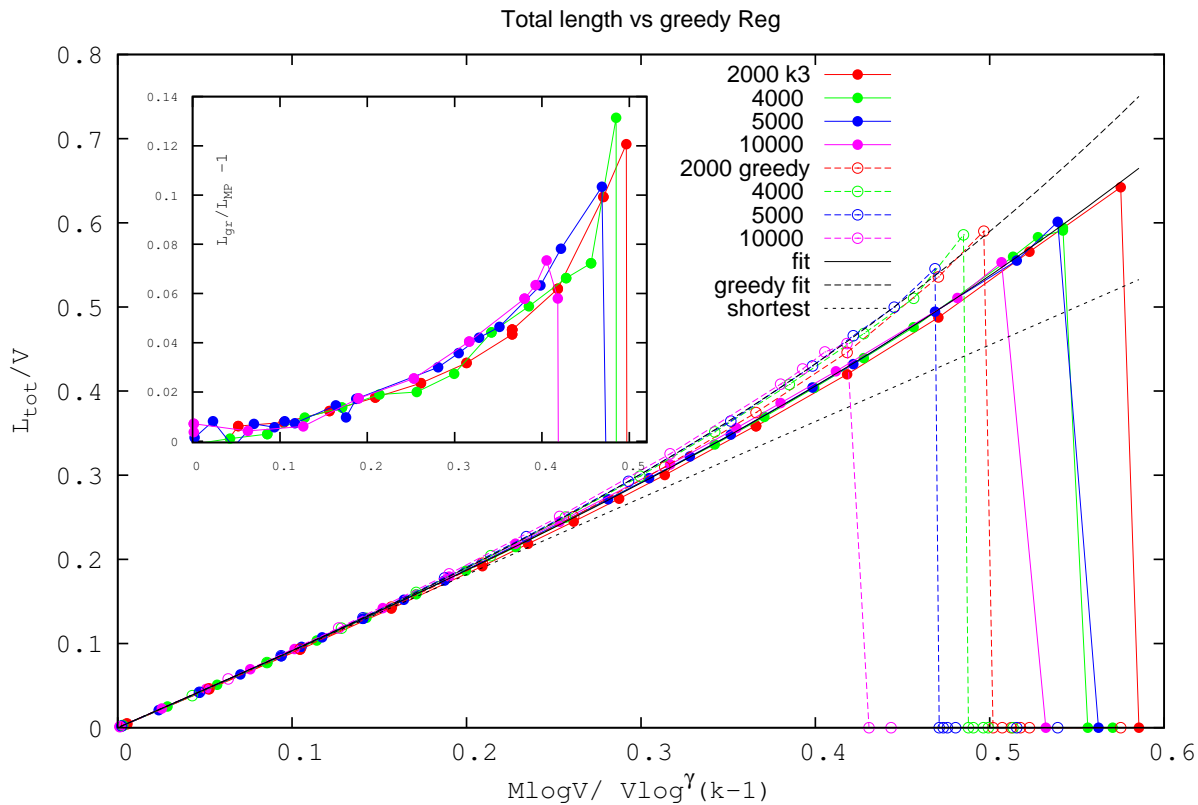}
\caption{Expected total normalized path length - greedy vs global optimization algorithms - Reg graphs. We compared results obtained by a greedy BFS algorithm and our global optimization for system size $V=2000,4000,5000,10000$,$\gamma=0.87$ and degree $k=3$. We can identify the sparse interval where the algorithms give similar results because communications are far apart and take the shortest paths. As the number of communications grow we observe an intermediate regime where global optimization performs better than BFS; and finally the dense regime where the greedy BFS algorithm fails to find a solution whereas the global optimization algorithm succeeds up to a critical  $M$ value. Cubic fits are also plotted (solid black line for global optimization, dashed line for the greedy BFS algorithm) whereas the dotted line represents the shortest path (not considering overlap) trivial solution, i.e. the sum of the $M$ shortest path lengths, which is linear in $x$. Vertical lines show the frustration points where no solution is found and the total path length is set to zero. Inset: Ratio $L_{greedy}/L_{MP} -1$ is plotted as a function of $x$. Notice the worse performance of  the greedy algorithm.}
\label{lgreedy}
\end{figure}

\begin{figure}[htbp]
\includegraphics[width=16cm]{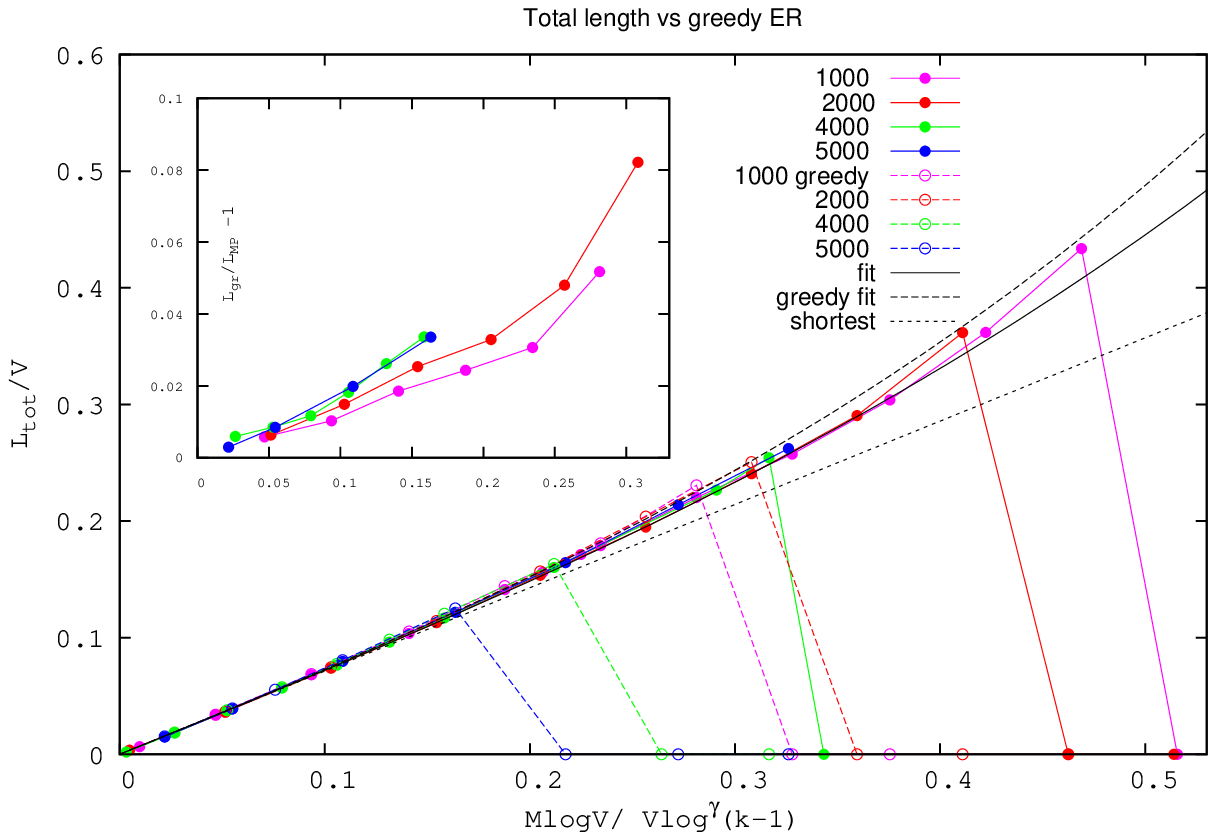}
\caption{Expected total normalized path length - greedy vs global optimization algorithms ER. System sizes $V=1000, 2000,4000,5000$, $\gamma=0.69$ and degree $\langle k \rangle=3$. Inset: the ratio $L_{greedy}/L_{MP} -1$ is plotted as a function of $x$. Notice the worse performance of  the greedy.}
\label{erlgreedy}
\end{figure}

\begin{figure}[htbp]
\includegraphics[width=16cm]{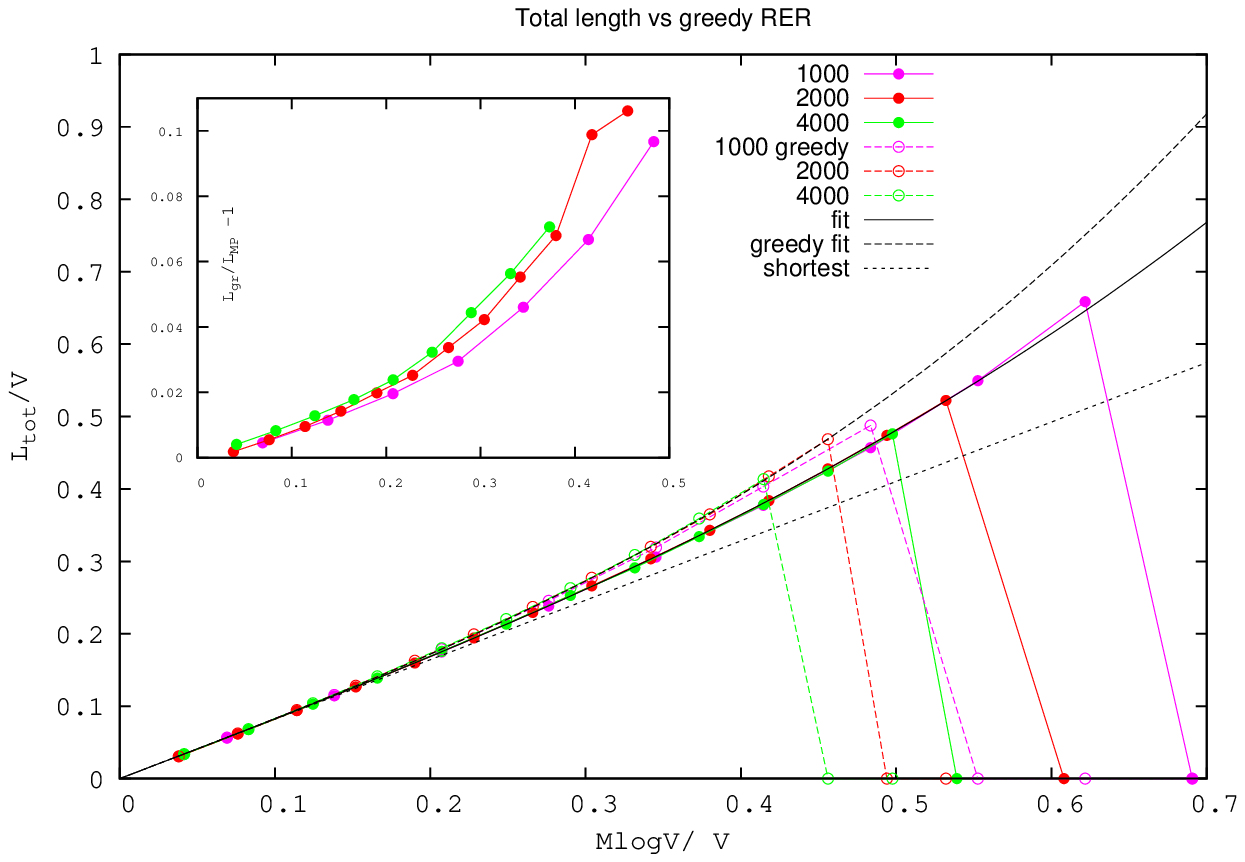}
\caption{Expected total normalized path length - greedy vs global optimization algorithms RER. System sizes $V=1000, 2000,4000$ and degree $\langle k \rangle=4$. Inset: the ratio $L_{greedy}/L_{MP} -1$ is plotted as a function of $x$. Notice the worse performance of  the greedy.}
\label{rerlgreedy}
\end{figure}

We found that our MP algorithm outperforms the greedy BFS both in finding a better solution (smaller $L_{tot}$) and in reaching higher values of the frustration transitions.
In figures ~\ref{lgreedy},  ~\ref{erlgreedy}  and  ~\ref{rerlgreedy}  we plotted the expected normalized total length for both greedy and MP algorithms, for the three graph types, fixed connectivity and different system sizes. We focused on the case $\langle k \rangle=3$ for Reg and ER and $\langle k \rangle=4$ for RER because of its lower complexity compared to higher $k$; nevertheless, simulations for different $k$ values indeed agree with the suggested scaling and exhibit the same behavior. Initially, in the $x$ range where solutions exist the greedy algorithm gives the same total length as the global algorithm up to a certain value of $M$ (and $x$). The explanation is that in this interval the graph is sparse, communications \emph{typically} do not interact and shortest paths can be selected. This also shows that for a small number of paths the global procedure reduces to act similarly to the greedy algorithm does, e.g. when rerouting is required it involves only two paths, the optimal solution will adopt the shortest path for one and will reroute the second. When $M$ increases, we see that the global optimization algorithm outperforms the greedy approach in both finding the optimal solution and in achieving a higher frustration threshold. In the regime where the global algorithm gives a better solution (i.e. shorter total length) we see that it is more efficient to globally reroute paths rather than taking the shortest path of selected paths and adapt the other paths. This means that the optimal solution is not a simple superposition of the first $n$-shortest path of the $M$ communications, but is a more complex solution.


\begin{figure}[htbp]
\begin{center}
\includegraphics[width=16cm]{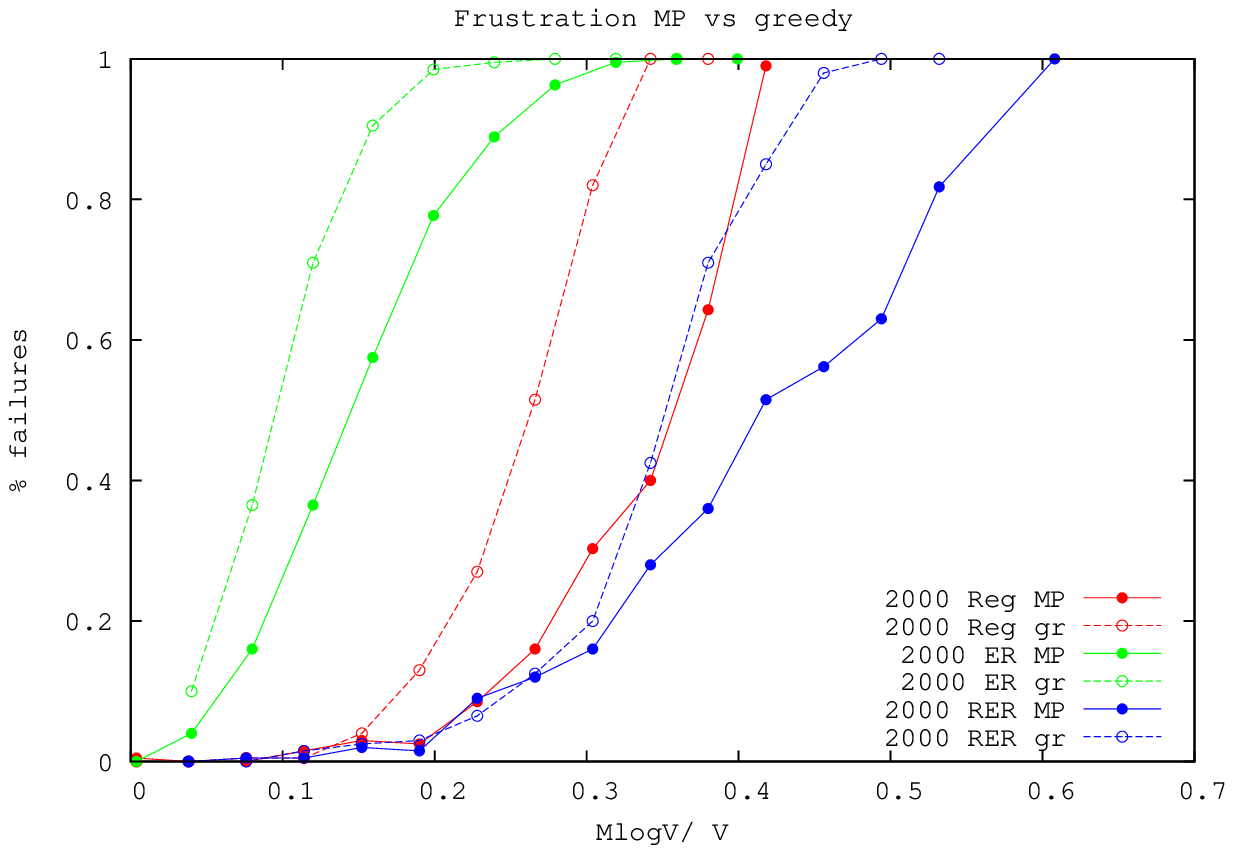}
\caption{Failure rate for greedy and MP algorithms. We plotted for $V=2000$ and for Reg and ER of degree $\langle k \rangle=3$ and RER of $\langle k \rangle=4$, the failure rate as a function of $\frac{M \log V}{V }$. We can notice that the greedy algorithm fails to find solution earlier than MP for all types of graph considered. ER reaches frustration sooner due to less path choices, whereas RER has higher frustration threshold because of the higher connectivity. The MP data shown are results of averages calculated over a smaller number of instances than for the greedy algorithm, hence the lines connecting them are less smooth.}\label{failure}
\end{center}
\end{figure}

Figure~\ref{failure} shows the failure ratio defined as the number of unsuccessful instances (for which a solution is not found) over the total number of realizations as a function of the scaling variable $x$. We notice that the greedy algorithm reaches the frustration point (as a function of $x$) earlier than the corresponding global MP algorithm, regardless the system sizes or graph type.

This shows that, if a solution exists, a global management of the entire set of communications is required in order to find an optimal solution. Whereas if each communication acts selfishly, seeking the corresponding shortest path, unsolvable overlaps between communications emerge at lower $x$ values. Both algorithms show an increased failure rate as the system size increases, presumably due to the unscaled limit on the number of iterations allowed and possibly inherent finite-size effects.

\subsection{A posteriori statistics: maximum cluster size and degree distribution (regular graph case).}
To better understand the optimization process and characterize the solutions obtained we carried out a statistical analysis of the solution a posteriori. Given the clearer statistical interpretation of the results obtained (due to the limited number of possible connectivity values and their evolution, and the higher frustration threshold for a given connectivity), we chose to study regular graphs. In this case one can gain more insights into the type of routes formed and the reduced effective graphs that emerge for any number of communications. By a posteriori we mean that once a solution was found, by an MP or greedy algorithm, we removed from the graph $\mathcal{G}$ all nodes and edges taking part in the paths and then calculated statistical properties of the remaining graph $\mathcal{G}'$. In particular, we calculated the maximum cluster size and the degree distribution.

\begin{figure}[htbp]
\begin{center}
\includegraphics[width=14cm]{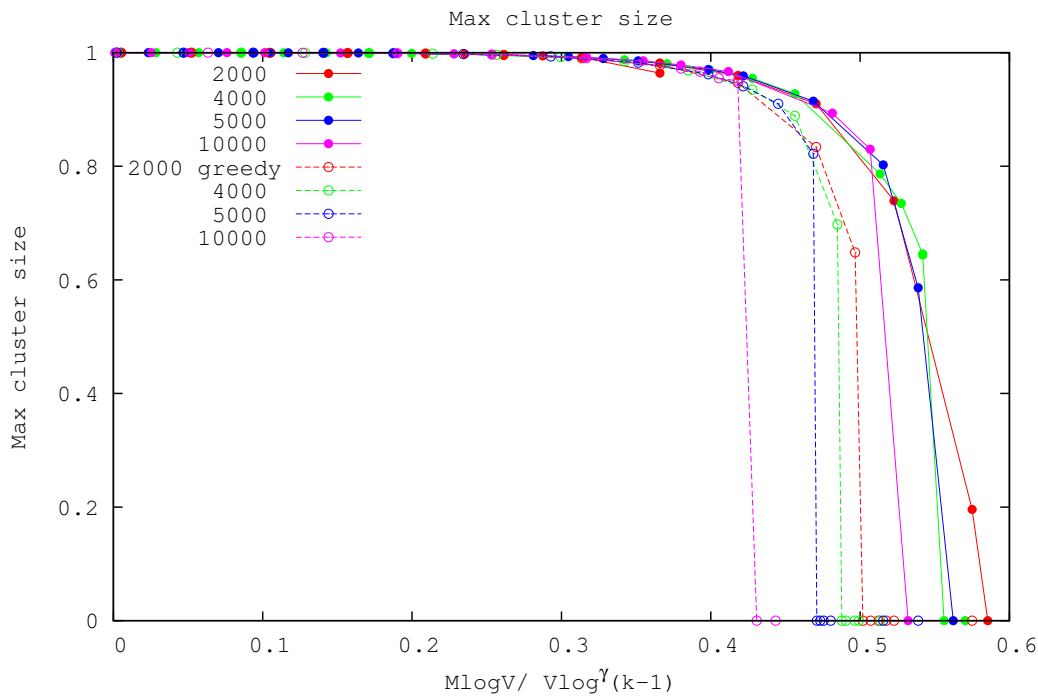}
\caption{Maximum cluster size greedy vs MP algorithms. The maximum cluster size, i.e. the ratio of giant component size normalized with respect to $V$, is plotted as a function of the scaling variable $x$. We see that in both cases frustration is reached for giant component values above $60 \%$; the lower values obtained for smaller graphs result from very few biased successful instances where source-destination pairs $(S,R)$ are selected from different clusters. }
\label{clustersize}
\end{center}
\end{figure}

The existence of a solution to a given set of communications is strictly related to the connectivity of the graph. Each time a solution for a subset of communications is found, edges and nodes involved in the solution paths are effectively removed; and properties of the reduced graph provide information on its ability to accommodate more source-destination pairs and the efficiency of the obtained solution in making use of the topology. Figure~\ref{clustersize} shows the max cluster sizes ratio of $\mathcal{G'}$  as a function of the scaling variable $x$.
 This quantity is defined as the ratio between the number of nodes in the maximum connected cluster over the number of nodes in the same graph $\mathcal{G'}$, the graph obtained after edge and node removal of the obtained solution paths. For both the greedy and the global MP algorithm we see an abrupt step change at some $x$ value, between a graph that has a giant connected component and a situation where no solution exists, such that we set the ratio to zero by convention. Moreover, this drop is more abrupt
 and occurs for smaller $x$ values in the case of the greedy algorithm. This means that the greedy procedure does not distribute paths evenly on the graph and creates small disconnected clusters; the greedy algorithm is therefore more sensitive to small changes in connectivity compared to the global MP algorithm, for which the drop is more gradual at first and occurs at higher $x$ values. This reconfirms the previous results that the greedy behavior is fragile and sensitive to the position of the communication pairs and the order in which they are selected.

\begin{figure}[htbp]
\begin{center}
\includegraphics[width=12cm]{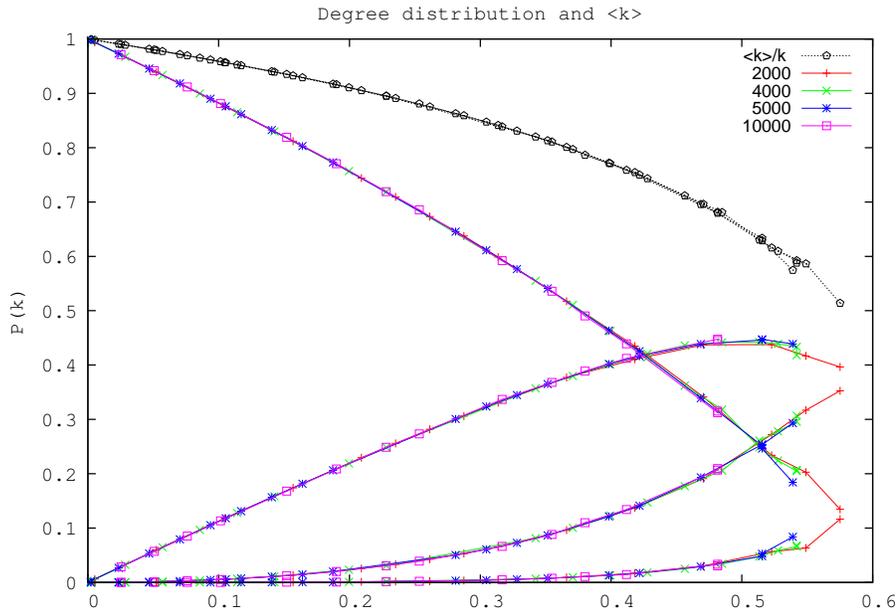}
\caption{Degree distribution and $\langle k \rangle$. We plotted $P(k)$ for $k=0,\dots,3$ as a function of $x=M \log V/V\log^{\gamma}(k-1)$ for a $3-$regular graph and system size $V=2000,4000,5000,10000$. From top to bottom: the curve at the top is $P(3)$ while the one at the bottom represents $P(0)$. The black line represents $\langle k \rangle /3$ for the different system sizes.}
\label{degree}
\end{center}
\end{figure}

We evaluate the a posteriori degree distribution $P(k)$ by calculating the connectivities $k_{i} \,\forall i \in\mathcal{G}'$ for the different $k$ values, and from these derive the average degree $\langle k \rangle$ as a function of the scaling variable $x$. Results shown in Figure~\ref{degree} for different system size and $k=3$ show consistent trends; starting from a $3-$regular graph we end up, close to the frustration transition point and after the node and edge removal, with about $20 \%$ of the nodes with $k=3$, whereas $\sim 40 \%$ have degree $k=2$ and $\sim 30 \%$ have $k=1$. The decay of $\langle k \rangle /k$ is also plotted for the same process. Also here we see a good data collapse (the different curves can only be distinguished close to frustration).

From graph theory~\cite{barabasi02} we know that when $\langle k\rangle \sim 1$ the graph is likely disconnected,
at least to two giant components; the numerical results show that frustration is reached when $\langle k \rangle/k$ has value $\sim 50-60 \%$, which corresponds to an average degree $\langle k \rangle \sim 1.5-1.8$, still higher than the connectivity threshold $1$. This can be explained by the fact that tighter constraints on edge availability are imposed in the case of the NDP problem, resulting in frustration even before the graph disconnects (i.e., disconnection is sufficient but not necessary for frustration). Indeed in our model it is insufficient to have just a good number of available links, but they should also constitute clusters of connected links to accommodate new communication paths. Hence the average connectivity value observed at the frustration point of $\langle k \rangle >1$.

\subsection{Path length and stretch distribution.}
\begin{figure}[htbp]
\begin{center}
\includegraphics[width=7cm]{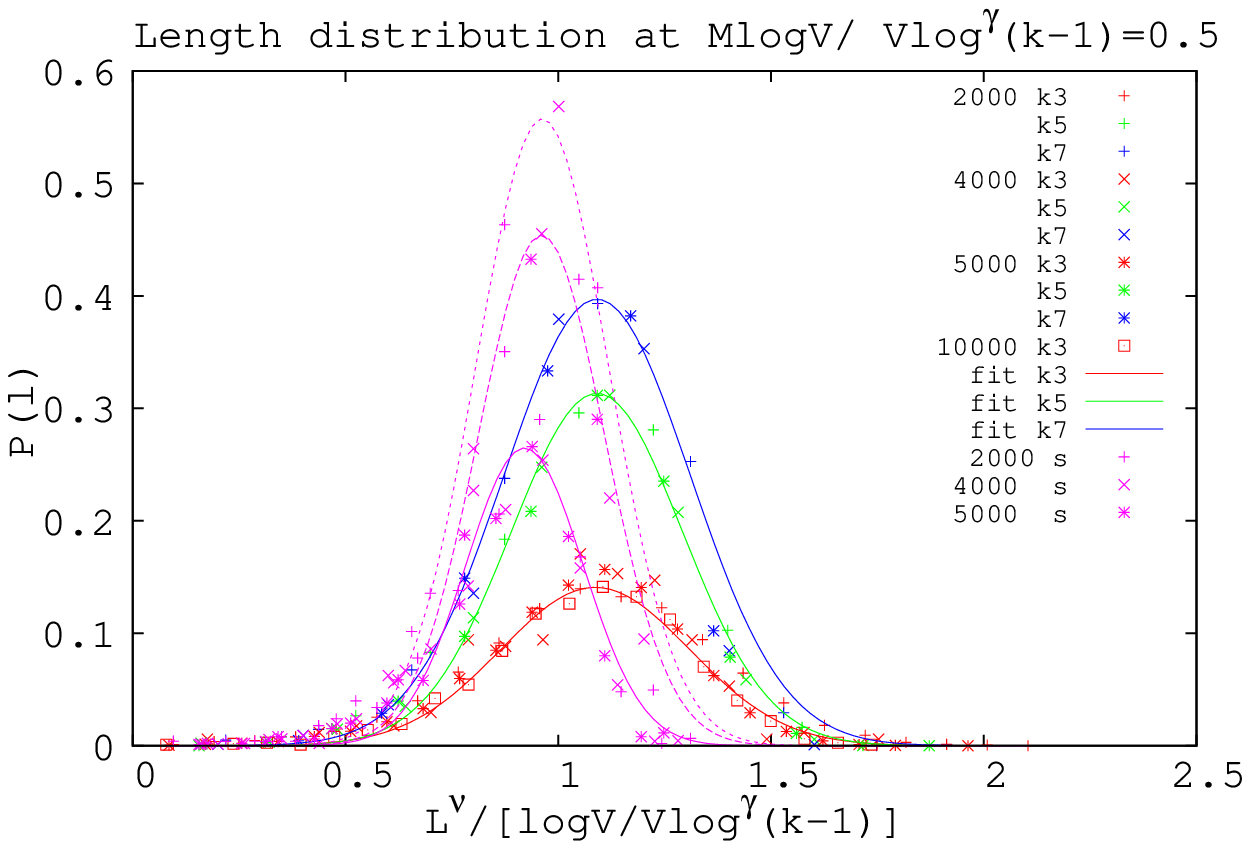}
\includegraphics[width=7cm]{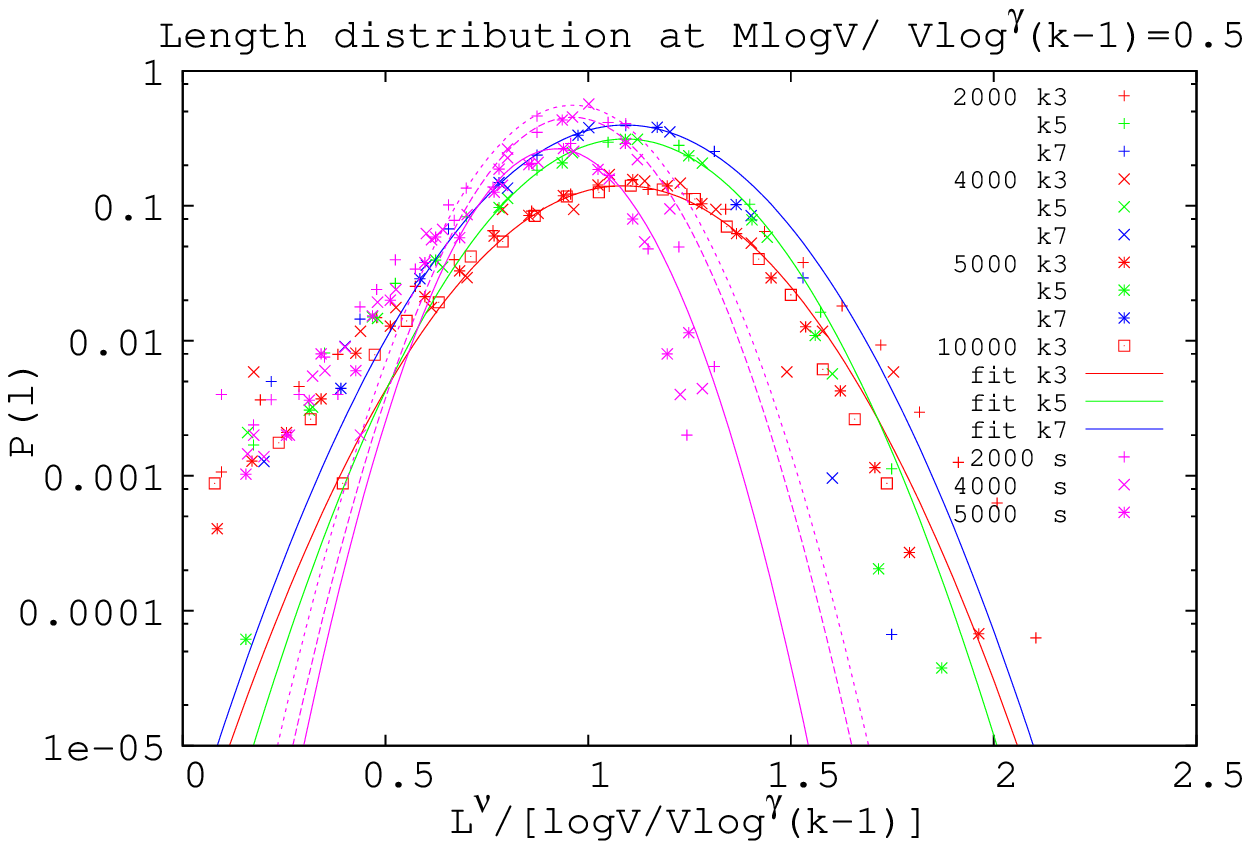}
\caption{Path length distribution. Left - Path lengths $L^{\nu}$  in the dense interval $M \log V/\log^{\gamma}(k-1)\sim 0.5$ are normalized and plotted for $k=3,5,7$ graphs against the corresponding shortest paths (violet curves). We see a more broadly spread distributions with higher averages for all graphs with respect to the shortest paths, signaling the path rerouting due to the MP optimization. Right - path lengths are plotted in log scale to highlight the left fat tails where the shortest paths are similar irrespective of connectivity, whereas other paths are rerouted by the MP algorithm and are considerably longer, almost by a factor of two (right tails). }
\label{ldistribution}
\end{center}
\end{figure}

Another interesting quantity to consider is the path length distribution close to the critical threshold, and its comparison with the shortest path distribution. Using the rescaled variable $\frac{L}{\log V/\log^{\gamma}(k-1)} $, where $L$ is the length per communication, we present in figure \ref{ldistribution} the distribution obtained for different system sizes. We see a good data collapse for graphs of different system sizes and connectivities to a Gaussian-like distribution with left fat tails, as confirmed by the log-plot on the right panel. This can be explained by the fact that the shorter of the $M$ shortest paths are less likely to be rerouted.
A graph with a high number of communications
exhibits a path length distribution with higher length averages (with respect to the shortest path distribution) as well as higher variances because solution path lengths are more broadly spread. We notice that the left tails are similar for all connectivity values whereas the right tails are broader for lower connectivities close to the frustration point. This can be explained by the fact that short paths are less likely to be rerouted and occur in roughly the same proportion in graphs of different connectivities; hence the similarity in the fat left tails. Regarding the right tails - many paths are rerouted through longer routes by the MP-algorithm, but graphs with higher degree allow for more communications with shorter routes due to the higher routing flexibility they offer.

\begin{figure}[htbp]
\begin{center}
\includegraphics[width=7cm]{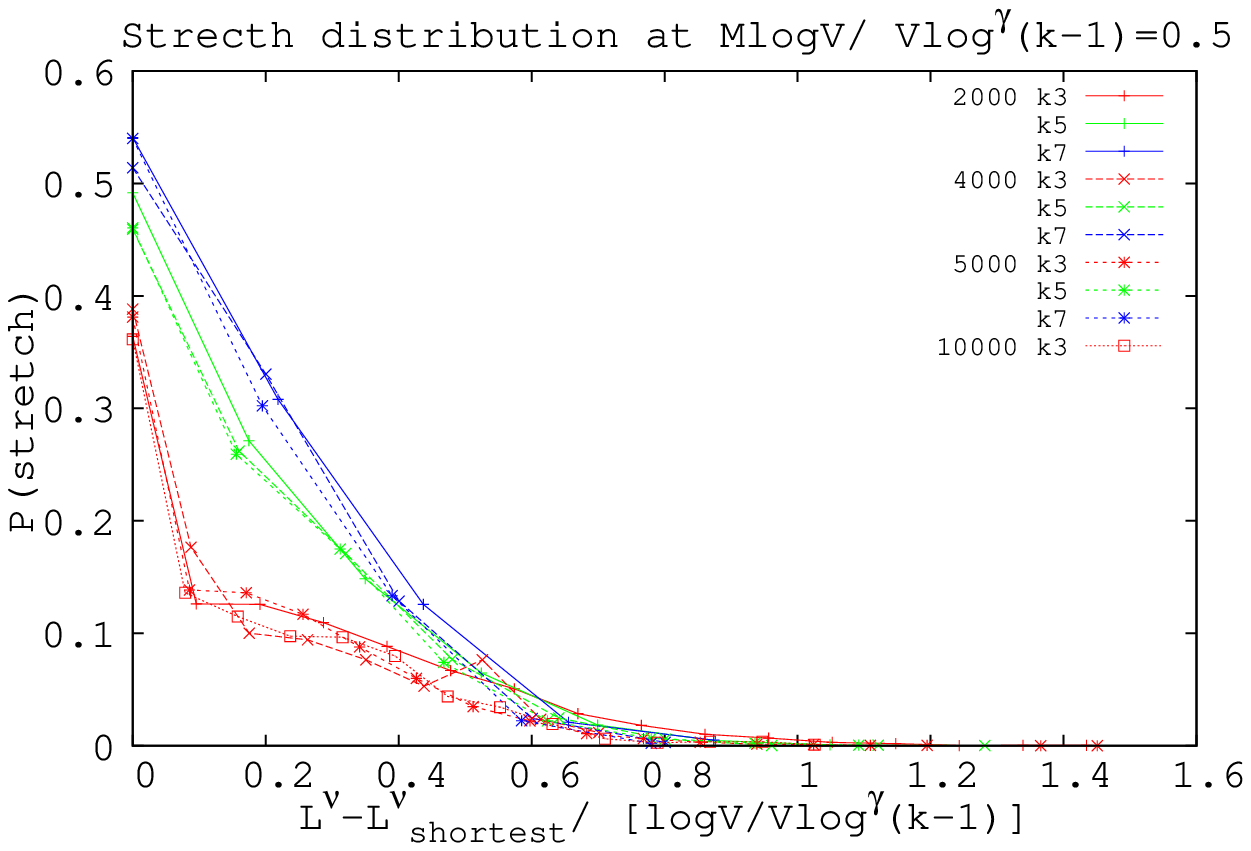}
\includegraphics[width=7cm]{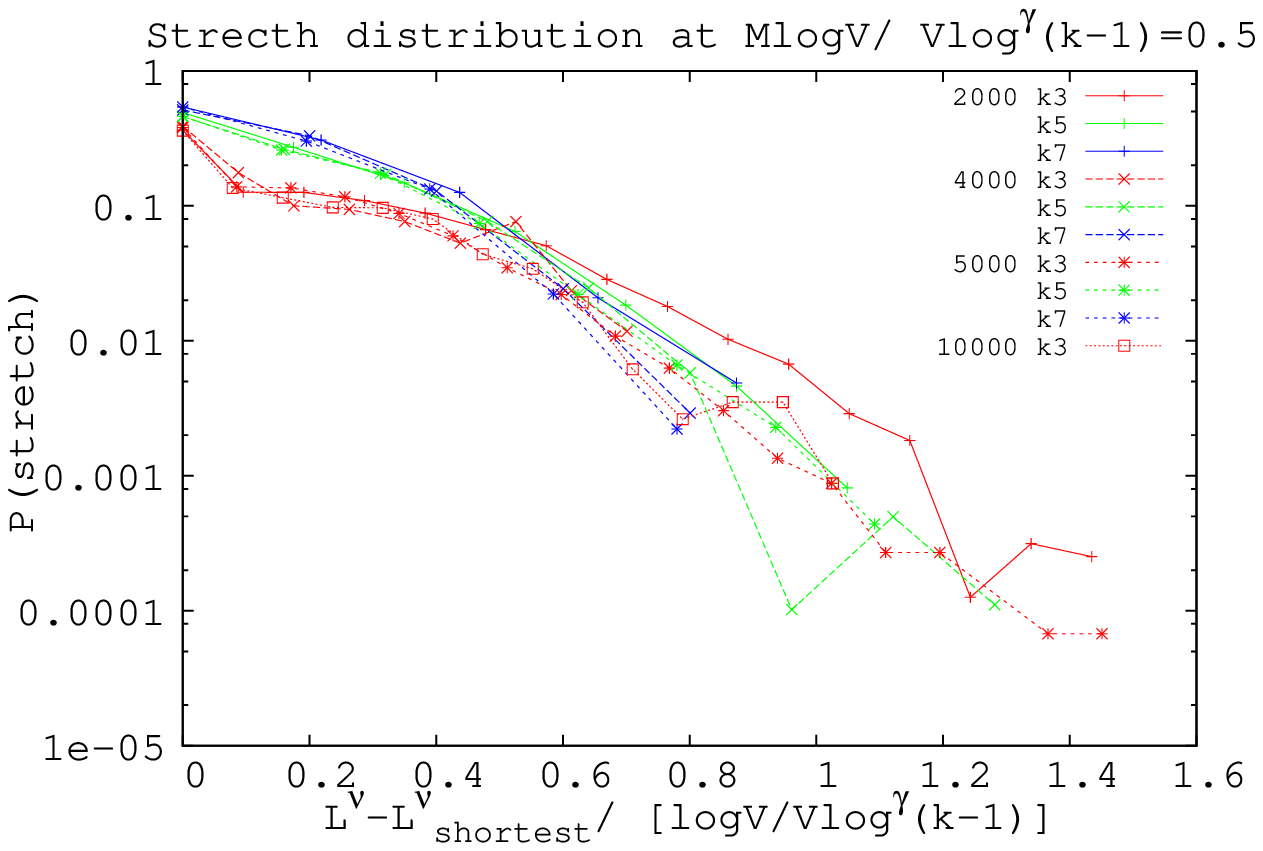}
\caption{Stretch distribution. Left - The difference $L^{\nu}- L^{\nu}_{shortest}$ is normalized and plotted for each communication in the dense regime for degrees $k=3,5,7$. For $k=3$ we have $40 \%$ of communications correspond to the shortest paths while others take routes which are long compared to the average path length $\log V/\log^{\gamma}(k-1)$. In the case of higher connectivities $k=5,7$ there is a higher fraction of paths with the same length as the shortest path due to the routing flexibility offered. Right - The same data is plotted on a log scale to highlight the tails behavior. }
\label{stretch}
\end{center}
\end{figure}

Figure \ref{stretch} shows the stretch, defined as the difference between the shortest path and the path length obtained through MP optimization, for $M$ close to frustration ($\frac{M\log V}{V \log^{\gamma}(k-1)} \sim 0.5$) for different system sizes.
We can see that for graphs of degree $k=3$ only $34-38\, \%$ of the communications follow the shortest path, all other communications are routed through longer paths. A higher fraction of shortest-path communications is found for higher connectivity graphs, presumably due to the higher routing flexibility they offer. Looking at the tails we can see that there is a non-negligible fraction of paths that stretch considerably compared to the average shortest path length.

\section{Conclusion}
\label{sec:conclusion}
We studied the shortest node-disjoint path problem on regular, ER and RER random graphs using message-passing cavity equations. We found that the suggested MP algorithm outperforms the greedy breadth-first search approach both in finding better solutions (shorter total path length) and in finding solutions for higher values of $M$. This shows that a  global strategy is needed to optimally route paths which do not overlap at nodes but also have minimal path lengths. We found a scaling rule for the total length that goes as a cubic function of the occupancy ratio $\frac{M\log V}{V \log^{\gamma}(k-1)}$, with $\gamma$ varying with the graph topology. This behavior resembles the shortest path length for small $M$ but increases faster than linearly for a higher number of paths.

We also studied statistical properties of physical observables a posteriori in the case of regular graphs. We found good data collapses for regular graphs of different system sizes and connectivities for quantities such as maximum cluster size, degree distribution and, length and stretch distributions.

We believe this approach is theoretically interesting due to its relevance to hard combinatorial complexity problems but also offers a new direction for solving important practical routing problem in communication, in particular in optical and wireless ad-hoc networks and VLSI design. This study offers the first step for realizing the potential in this new direction.

\section*{Acknowledgement}
This work is supported by the Marie Curie Training Network NETADIS (FP7, grant 290038), the EU FET FP7 project STAMINA (FP7-265496) and the Royal Society Exchange Grant IE110151. This work is partially supported by Research Grants Council of Hong Kong (605010 and 604512).

\section*{References.}

\bibliography{bibliography}{}
\bibliographystyle{unsrt}

\end{document}